\documentclass[a4paper,12pt]{article}
\usepackage{amssymb}
\usepackage{amsmath}
\usepackage{fullpage}
\usepackage{boxedminipage}
\usepackage{listings}
\usepackage{minitoc}
\usepackage{graphicx}
\makeatletter \@addtoreset{equation}{section} \makeatother

\begin{document}
\input epsf

\begin{titlepage}

    \thispagestyle{empty}
    \begin{flushright}
        \hfill{CERN-PH-TH/2007-140} \\
    \end{flushright}

    \vspace{5pt}
    \begin{center}
        { \Huge{\textbf{Black Hole Attractors\\\vspace{10pt}in Extended Supergravity}}}\vspace{25pt}
        \vspace{55pt}

        {\textbf{Sergio Ferrara}$^{\diamondsuit\clubsuit\flat}$ \textbf{and\ Alessio Marrani}$^{\heartsuit\clubsuit}$}

        \vspace{15pt}

        {$\diamondsuit$ \it Physics Department,Theory Unit, CERN, \\
        CH 1211, Geneva 23, Switzerland\\
        \texttt{sergio.ferrara@cern.ch}}

        \vspace{10pt}

        {$\clubsuit$ \it INFN - Laboratori Nazionali di Frascati, \\
        Via Enrico Fermi 40,00044 Frascati, Italy\\
        \texttt{marrani@lnf.infn.it}}

        \vspace{10pt}

         {$\flat$ \it Department of Physics and Astronomy,\\
        University of California, Los Angeles, CA USA}

         \vspace{10pt}

        {$\heartsuit$ \it Museo Storico della Fisica e\\
        Centro Studi e Ricerche ``Enrico Fermi"\\
        Via Panisperna 89A, 00184 Roma, Italy}

        \vspace{50pt}
        \noindent \textit{Contribution to the Proceedings of PASCOS $2007$,\\$13$th International Symposium on Particles, Strings and Cosmology,\\2--7 July 2007, Imperial College, London, UK,\\to be published online by the American Institute of Physics}
\end{center}


\begin{abstract}
We review some aspects of the attractor mechanism for extremal black
holes of (not necessarily supersymmetric) theories coupling Einstein
gravity to
scalars and Maxwell vector fields. Thence, we consider $\mathcal{N}=2$ and $%
\mathcal{N}=8$, $d=4$ supergravities, reporting some recent advances
on the moduli spaces associated to BPS and non-BPS attractor
solutions supported by charge orbits with non-compact stabilizers.
\end{abstract}

\end{titlepage}
\newpage
The so-called \textit{attractor mechanism} was first considered in the
framework of $\mathcal{N}=2$, $d=4$ ungauged supergravity coupled to $n_{V}$
vector multiplets \cite{FKS}-\nocite{Strom,FK1,FK2}\cite{FGK}. It concerns
the stabilization of the scalar fields $\phi ^{i}$ ($i=1,...,n_{V}$) of the
theory near the event horizon of an extremal, static, spherically symmetric
and asymptotically flat black hole (BH) \cite{black}. An extremal BH can be
defined to have vanishing temperature ($T=0$), and thus it is
thermodynamically stable. The asymptotical behavior of the scalars $\phi ^{i}
$ is defined by the limits
\begin{eqnarray}
lim_{r\rightarrow \infty }\phi ^{i}\left( r\right) &=&\phi _{\infty }^{i}\in
\mathcal{M};  \label{asymptotical-limit} \\
lim_{r\rightarrow r_{H}}\phi ^{i}\left( r\right) &=&\phi _{H}^{i}\left(
q,p\right) ,  \label{horizon-limit}
\end{eqnarray}
where $\mathcal{M}$ is the scalar manifold, $r_{H}$ is the radial coordinate
of the event horizon, and $\left( q,p\right) $ denotes the set $\left\{
q_{\Lambda },p^{\Lambda }\right\} $ of the electric and magnetic charges of
the BH ($\Lambda =0,1,...,n_{V}$), which are conserved due to the overall $%
\left( U(1)\right) ^{n_{V}+1}$ gauge-invariance of the considered theory.
The dynamical flow determining the radial evolution of the scalars $\phi
^{i}\left( r\right) $ between the above two asymptotical limits is
non-singular near the horizon, provided that
\begin{equation}
\left. \frac{\partial V_{BH}\left( \phi ,q,p\right) }{\partial \phi ^{i}}%
\right| _{\phi ^{j}=\phi _{H}^{j}}=0,  \label{extremum-cond}
\end{equation}
where $V_{BH}$ is a certain positive definite, charge-dependent function in $%
\mathcal{M}$, named \textit{BH effective potential }\cite{FGK}. The
condition (\ref{extremum-cond}) determines the so-called
\textit{attractor equations}, whose solutions are the purely
charge-dependent, stabilized horizon configurations $\phi
_{H}^{i}\left( q,p\right) $ in the r.h.s. of Eq.
(\ref{horizon-limit}). By using the Bekenstein-Hawking entropy-area
formula \cite{BH1,FGK}, the classical BH entropy reads
\begin{equation}
\mathcal{S}\left( q,p\right) =\frac{A_{H}}{4}=\pi V_{BH}\left( \phi
_{H}\left( q,p\right) ,q,p\right) ,
\end{equation}
where $A_{H}$ is the area of the BH event horizon.

The horizon geometry of extremal, asymptotically flat BHs in $\mathcal{N}=2$%
, $d=4$ supergravity is a maximally supersymmetric $\mathcal{N}{=2}$
background, namely the Bertotti-Robinson (BR) $AdS_{2}\times S^{2}$
BH metric \cite{BR1,Lowe-Strominger1}, which in turn is a particular
case of the extremal $p$-brane horizon geometry $AdS_{p+2}\times
S^{d-p-2}$ \cite {Gibbons-Townsend1}.
\begin{figure}[p]
\centerline{ \epsfxsize 4in\epsfbox{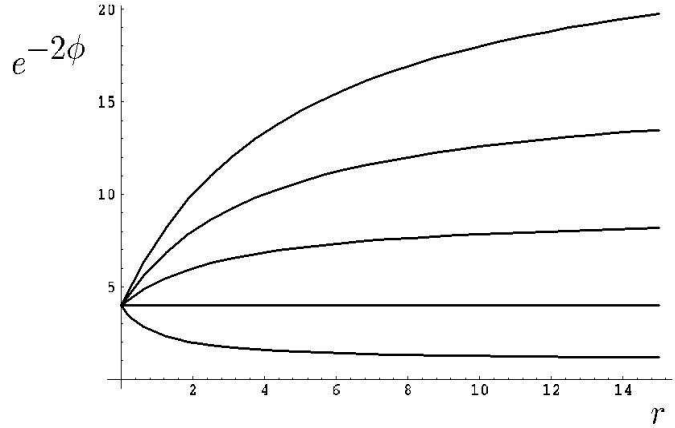}}
\caption{\textbf{Realization of the attractor mechanism in the $\frac{1}{2}$%
-BPS\ dilatonic BH \protect\cite{FK1,FK2,black}}. Independently on the set
of asymptotical ($r\rightarrow \infty $) scalar configurations, the
near-horizon evolution of the dilatonic function $e^{-2\protect\phi }$
converges towards a fixed \textit{attractor} value, which is purely
dependent on the (ratio of the) quantized conserved charges of the BH.}
\end{figure}
\begin{figure}[p]
\centerline{ \epsfxsize 4in\epsfbox{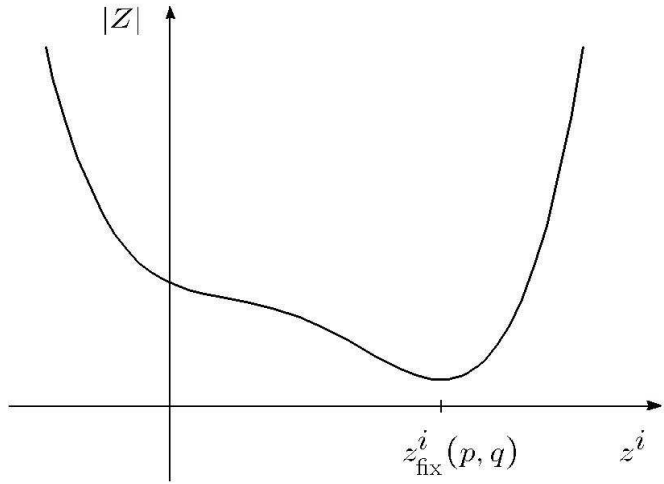}}
\caption{\textbf{Minimization of the absolute value of the ``central
charge'' function $Z$ in $\mathcal{M}$.} In the picture
$z_{fix}^{i}\left( p,q\right) $ stands ~for ~the
~\textit{attractor}, purely charge-dependent
value of the scalars at the event horizon of the considered $\frac{1}{2}$%
-BPS\ extremal BH. The attractor mechanism fixes the extrema of the
central
charge to correspond to the discrete \textit{fixed} points of the \textit{%
attractor variety} \protect\cite{Moore1} $\mathcal{M}$. Of course, the
dependence of the central charge on scalars is shown for a given supporting
BH charge configuration.}
\end{figure}

The first class of attractors to be studied was the
$\frac{1}{2}$-BPS one, which preserves $4$ supersymmetries out of
the $8$ pertaining to the asymptotical $\mathcal{N}=2$, $d=4$
Poincar\'{e} superalgebra. Examples of such attractors are given by
Figures 1 and 2. Recently, many important advances have been
performed in the study of extremal BH attractors, mainly concerning
new classes of attractor configurations, corresponding to non-BPS,
non-supersymmetric horizon geometries \cite{Sen-old1}--\nocite
{GIJT,Sen-old2,K1,TT,G,GJMT,Ebra1,K2,Ira1,Tom,BFM,AoB-book,FKlast,Ebra2,BFGM1,rotating-attr,K3,Misra1,Lust2,BFMY,CdWMa,DFT07-1,BFM-SIGRAV06,Cer-Dal,ADFT-2,Saraikin-Vafa-1,Ferrara-Marrani-1,TT2,ADOT-1,Ferrara-Marrani-2,CCDOP,Misra2,CFM1}\cite{BMOS-1}.

For asymptotically flat extremal BHs $V_{BH}$ is given in terms of the
scalar-dependent, complex symmetric matrix $\mathcal{N}%
_{\Lambda \Sigma }\left( \phi \right) $ (with $Im\mathcal{N}_{\Lambda \Sigma
}$ negative definite), determining the couplings of the Maxwell field
strength terms $\mathcal{F}^{2}$ and $\mathcal{F}\widetilde{\mathcal{F}}$ in
the Lagrangian density, and of the electric and magnetic BH charges \cite
{FGK}:
\begin{equation}
V_{BH}\left( \phi ,q,p\right) =-\frac{1}{2}\left( q_{\Lambda }-\mathcal{N}%
_{\Lambda \Sigma }p^{\Sigma }\right) \left( Im\mathcal{N}\right) ^{-1\mid
\Lambda \Delta }\left( q_{\Delta }-\overline{\mathcal{N}}_{\Delta \Gamma
}p^{\Gamma }\right) .
\end{equation}
Such a formula is valid for any (not necessarily supersymmetric) theory
coupling Einstein gravity to scalars and Maxwell vector fields, whose
Lagrangian density in general has the form
\begin{equation}
\begin{array}{l}
\frac{\mathcal{L}}{\sqrt{-g}}=-\frac{1}{2}R-g_{ab}\left( \partial _{\mu
}\phi ^{a}\right) \left( \partial _{\nu }\phi ^{b}\right) G^{\mu \nu }+ \\
\\
+\left( Im\mathcal{N}_{\Lambda \Sigma }\right) \mathcal{F}_{\mu \nu
}^{\Lambda }\mathcal{F}^{\Sigma \mid \mu \nu }+\frac{1}{2\sqrt{-g}}\left( Re%
\mathcal{N}_{\Lambda \Sigma }\right) \epsilon ^{\mu \nu \rho \lambda }%
\mathcal{F}_{\mu \nu }^{\Lambda }\mathcal{F}_{\rho \lambda }^{\Sigma }+...,
\end{array}
\end{equation}
where $g_{ab}$ is the metric of the scalar manifold and $G_{\mu \nu }$ is
the space-time metric.

An equivalent (but manifestly duality-covariant) expression reads \cite{FGK}
\begin{equation}
V_{BH}\left( \phi ,q,p\right) =-\frac{1}{2}Q^{T}M\left( \mathcal{N}\right) Q,
\end{equation}
where $Q^{T}$ is the $1 \times \left( 2n_{V}+2\right)$ vector
$\left(
p^{\Lambda },q_{\Lambda }\right) $ of the BH charges, and $M\left( \mathcal{N%
}\right) $ is the symplectic $\left( 2n_{V}+2\right) \times \left(
2n_{V}+2\right) $ real, negative definite symmetric matrix
\begin{equation}
\begin{array}{l}
M=R^{T}M_{D}R,~~R\equiv \left(
\begin{array}{cc}
I & 0 \\
-Re\mathcal{N} & I
\end{array}
\right) ,~~M_{D}\equiv \left(
\begin{array}{cc}
Im\mathcal{N} & 0 \\
0 & \left( Im\mathcal{N}\right) ^{-1}
\end{array}
\right) , \\
\\
M\Omega M=\Omega ,~\Omega \equiv \left(
\begin{array}{cc}
0 & -I \\
I & 0
\end{array}
\right) .
\end{array}
\end{equation}

In $\mathcal{N}=2$, $d=4$ supergravity the scalar manifold is endowed with
the so-called special K\"{a}hler geometry (see \textit{e.g.} \cite
{CDF-review}) ; its metric $g_{i\overline{j}}=\partial _{i}\overline{%
\partial }_{\overline{j}}K$ ($K$ being the K\"{a}hler potential) and $%
\mathcal{N}_{\Lambda \Sigma }$ are respectively given by the formul\ae :
\begin{equation}
\begin{array}{l}
g_{i\overline{j}}=-ie^{K}\left[ \left( \overline{D}_{\overline{j}}\overline{X%
}^{\Lambda }\right) D_{i}F_{\Lambda }-\left( D_{i}X^{\Lambda }\right)
\overline{D}_{\overline{j}}\overline{F}_{\Lambda }\right] ; \\
\\
\mathcal{N}_{\Lambda \Sigma }=h_{I\Lambda }\left( f^{-1}\right) _{\Sigma
}^{I},~~f_{I}^{\Lambda }\equiv e^{K/2}\left( X^{\Lambda },\overline{D}_{%
\overline{i}}\overline{X}^{\Lambda }\right) ,~~h_{I\Lambda }\equiv
e^{K/2}\left( F_{\Lambda },\overline{D}_{\overline{i}}\overline{F}_{\Lambda
}\right) ,
\end{array}
\end{equation}
where $D_{i}$ denotes the K\"{a}hler-covariant derivative, and $\left(
X^{\Lambda },F_{\Lambda }\right) $ are the holomorphic sections of the Hodge
bundle over $\mathcal{M}$ ($F_{\Lambda }=\frac{\partial F\left( X\right) }{%
\partial X^{\Lambda }}$, whenever the holomorphic prepotential function $%
F\left( X\right) $ exists) (see \textit{e.g.} \cite {CDF-review} and
Refs. therein).

The symplectic-covariant formulation of the $\mathcal{N}=2$ special
K\"{a}hler geometry can be actually generalized to all extended ($\mathcal{N}=3,...,8$%
) $d=4 $ supergravities \cite{Andrianopoli:1996ve,FKlast,ADFT}. In such
theories, $V_{BH}$ can be expressed as
\begin{equation}
V_{BH}=\frac{1}{2}\left| Z_{AB}\right| ^{2}+\left| Z^{I}\right| ^{2},
\label{VBH-extended}
\end{equation}
where $Z_{AB}$ ($A,B=1,...,\mathcal{N}$) is the antisymmetric \textit{%
central charge matrix} and $Z^{I}$ are the so-called \textit{dressed} (or
\textit{matter}) \textit{charges}, respectively appearing in the
supersymmetry transformations of the gravitinos $\psi _{\mu A}$ and of the
other fermions $\lambda _{A}^{I}$ of the theory in the considered BH
background:
\begin{equation}
\begin{array}{l}
\left. \delta _{\varepsilon }\psi _{\mu A}\right| _{BH}\sim Z_{AB}\gamma
_{\mu }\varepsilon ^{B}; \\
\\
\left. \delta _{\varepsilon }\lambda _{A}^{I}\right| _{BH}\sim
Z^{I}\varepsilon _{A},
\end{array}
\end{equation}
where $\gamma _{\mu }$ are the $\gamma $-matrices and $\varepsilon $ is the
parameter of the supersymmetric transformation.

Let us consider the maximal $d=4$ supergravity, \textit{i.e.} $\mathcal{N}=8$
supergravity, based on the real $70$-dim. symmetric manifold $\frac{E_{7(7)}%
}{SU(8)}$ \cite{Cremmer:1979up}. In this case no matter multiplets
are coupled to the gravity one, thus Eq. (\ref{VBH-extended})
simplifies to ($A,B=1,...,8$)
\begin{equation}
V_{BH}=\frac{1}{2}\left| Z_{AB}\right| ^{2},
\end{equation}
with $Z_{AB}=L_{AB}^{\Lambda }\left( \phi \right) Q_{\Lambda }$, where $%
L\left( \phi \right) \in E_{7(7)}$ and $Q$ is the charge vector. Under a
transformation $h$ of the stabilizer $SU(8)$, the matrix $Z$ transforms as
\cite{Ferrara-Marrani-2}
\begin{equation}
Z\left( \phi ,Q\right) \longmapsto Z\left( \phi _{g},Q\right) =hZ\left( \phi
_{g},g^{-1}Q\right) \Longrightarrow V_{BH}\left( \phi ,Q\right)
=V_{BH}\left( \phi _{g},g^{-1}Q\right) .  \label{6Aug-1}
\end{equation}
By computing $V_{BH}$ at one of its critical points, one obtains a
completely charge-dependent expression:
\begin{equation}
\left. V_{BH}\right| _{\frac{\partial V_{BH}}{\partial \phi }=0}\equiv
V_{BH,cr.}\left( Q\right) =V_{BH,cr.}\left( g^{-1}Q\right) \sim \sqrt{\left|
\mathcal{J}_{4}\right| },
\end{equation}
$\mathcal{J}_{4}$ being the quartic Cartan-Cremmer-Julia invariant
of the fundamental representation $\mathbf{56}$ of $E_{7(7)}$ \cite
{Cremmer:1979up,Cartan}.

The local $SU(8)$ symmetry allows one to go to the so-called ``normal
frame'' \cite{Ferrara:1980ra}. In such a frame, $Z_{AB}$ and $\mathcal{J}%
_{4} $ respectively read as follows:
\begin{equation}
\begin{array}{l}
Z_{AB,normal}=skew-diag\left( \rho _{1},\rho _{2},\rho _{3},\rho _{4}\right)
e^{i\varphi /4}; \\
\\
\mathcal{J}_{4,normal}=\Big [(\rho _{1}+\rho _{2})^{2}-(\rho _{3}+\rho
_{4})^{2}\Big]\Big [(\rho _{1}-\rho _{2})^{2}-(\rho _{3}-\rho _{4})^{2}\Big]%
+8\rho _{1}\rho _{2}\rho _{3}\rho _{4}(cos\varphi -1),
\end{array}
\end{equation}
with $\rho _{i}\in R^{+}~\forall i=1,...,4$. Note that $Z_{AB,normal}$ has
an $\left( SU(2)\right) ^{4}$ symmetry.

From the analysis performed in \cite{FM,FG,FKlast}, the $\mathcal{N}=8$
attractor equations yield only 2 distinct classes of solutions with
non-vanishing entropy ($\frac{1}{8}$-BPS for $\mathcal{J}_{4}>0$, non-BPS
for $\mathcal{J}_{4}<0$):\medskip

\textbf{1. }$\frac{1}{8}$\textbf{-BPS:} $\rho _{1}=\rho _{\frac{1}{8}%
-BPS}\in R_{0}^{+}$ and all the others vanish, $\mathcal{J}_{4,normal,\frac{1%
}{8}-BPS}>0$, and
\begin{equation}
\mathcal{S}_{\frac{1}{8}-BPS}=\pi \sqrt{\mathcal{J}_{4,normal,\frac{1}{8}%
-BPS}}=\pi \rho _{1}^{2}.  \label{primera-1}
\end{equation}
The corresponding orbit of supporting BH charges in the $\mathbf{56}$ of $%
E_{7(7)}$ is $\mathcal{O}_{\frac{1}{8}-BPS}=\frac{E_{7(7)}}{E_{6(2)}}$.
Moreover, $Z_{AB,normal,\frac{1}{8}-BPS}$ has symmetry enhancement ($m.c.s.$
stands for \textit{maximal compact subgroup})
\begin{equation}
\left( SU(2)\right) ^{4}\longrightarrow SU(6)\otimes SU(2)=m.c.s.\left(
E_{6(2)}\right) .\smallskip
\end{equation}
Notice that $\varphi _{\frac{1}{8}-BPS}$ is actually undetermined.

\textbf{2. non-BPS:} all $\rho $s are equal to $\rho _{non-BPS}\in R_{0}^{+}$%
, $\varphi _{non-BPS}=\pi $, $\mathcal{J}_{4,normal,non-BPS}<0$, and \textbf{%
\ }
\begin{equation}
\mathcal{S}_{non-BPS}=\pi \sqrt{-\mathcal{J}_{4,normal,non-BPS}}=4\pi \rho
^{2}.  \label{segunda-1}
\end{equation}
The corresponding orbit of supporting BH charges in the $\mathbf{56}$ of $%
E_{7(7)}$ is $\mathcal{O}_{non-BPS}=\frac{E_{7(7)}}{E_{6(6)}}$. Furthermore,
$Z_{AB,normal,non-BPS}$ has symmetry enhancement
\begin{equation}
\left( SU(2)\right) ^{4}\longrightarrow USp(8)=m.c.s.\left( E_{6(6)}\right) .
\end{equation}

Thus, the symmetry of $Z_{AB,normal}$ gets enhanced at the particular points
of $\frac{E_{7(7)}}{SU(8)}$ given by the solutions of $\mathcal{N}=8$, $d=4$
attractor equations with non-vanishing $\mathcal{J}_{4}$. In general,
\textit{the invariance properties of the solutions to attractor Eqs. with }$%
\mathcal{J}_{4}\neq 0$ \textit{are given by the m.c.s. of the stabilizer of
the corresponding supporting BH charge orbit}.

The $70\times 70$ Hessian matrix of $V_{BH}$ at the $\frac{1}{8}$-BPS
critical points has rank $30$; its $30$ strictly positive and $40$ vanishing
eigenvalues respectively correspond to the $15$ vector multiplets and to the
$10$ hypermultiplets of the $\mathcal{N}=2$, $d=4$ spectrum obtained by
reducing $\mathcal{N}=8$ supergravity according to the following branching
of the $\mathbf{70}$ (four-fold antisymmetric) of $SU(8)$ \cite{ADF2}:
\begin{equation}
\begin{array}{l}
SU(8)\longrightarrow SU(6)\otimes SU(2); \\
\\
\mathbf{70}\longrightarrow \left[ \left( \mathbf{15},\mathbf{1}\right)
\oplus \left( \overline{\mathbf{15}},\mathbf{1}\right) \right] _{m\neq
0}\oplus \left( \mathbf{20},\mathbf{2}\right) _{m=0}.
\end{array}
\label{SU(8)-->SU(2)xSU(6)}
\end{equation}
On the other hand, at the non-BPS critical points the Hessian matrix
has rank $28$; such a splitting of the mass spectrum can be
interpreted according to the
following branching of the $\mathbf{70}$ of $SU(8)$ \cite{Ferrara-Marrani-1}%
:
\begin{equation}
\begin{array}{l}
SU(8)\longrightarrow USp(8); \\
\\
\mathbf{70}\longrightarrow \left( \mathbf{1}\oplus \mathbf{27}\right)
_{m\neq 0}\oplus \left( \mathbf{42}\right) _{m=0}.
\end{array}
\end{equation}

As shown in \cite{Ferrara-Marrani-2}, the massless modes of the critical
Hessian matrix actually correspond to flat directions of $V_{BH}$ itself.
This can be easily realized by noticing that the stabilizers of the charge
orbits are non-compact, so that
\begin{equation}
\begin{array}{l}
g_{Q}Q^{BPS}=Q^{BPS},~\forall g_{Q}\in E_{6(2)}; \\
\\
g_{Q}Q^{non-BPS}=Q^{non-BPS},~\forall g_{Q}\in E_{6(6)},
\end{array}
\end{equation}
and thus at the critical points (recall Eq. (\ref{6Aug-1}))
\begin{equation}
V_{BH}\left( \phi _{g_{Q}},g_{Q}^{-1}Q\right) =V_{BH}\left( \phi
_{g_{Q}},Q\right) =V_{BH}\left( \phi ,Q\right) .
\end{equation}
This implies that each of the two classes of $\mathcal{N}=8$, $d=4$ extremal
BH attractors with non-vanishing entropy has an associated moduli space:
\begin{equation}
\begin{array}{l}
BPS:\frac{E_{6(2)}}{SU(6)\otimes SU(2)}%
,~quaternionic~manifold~with~dim_{R}=40; \\
\\
non-BPS:\frac{E_{6(6)}}{USp(8)},~\mathcal{N}%
=8,~d=5~scalar~manifold~with~dim_{R}=42.
\end{array}
\end{equation}

The same reasoning, which is actually independent on the number $d$
of space-time dimensions and on $\mathcal{N}$, will apply to
\textit{all} theories of the kind considered above, whose scalar
manifold is an homogeneous (not necessarily symmetric) space, when
the stabilizer of the orbit of the attractor-supporting charge
vector $Q$ is non-compact \cite {Ferrara-Marrani-2}. For
$\mathcal{N}>2$ this will apply to both BPS and non-BPS critical
points (as shown above for $\mathcal{N}=8$, $d=4$).
However, for $\mathcal{N}=2$, $d=4$ the stabilizer of the ($\frac{1}{2}$%
-)BPS orbit is compact, and no flat directions will occur (apart from
hypermultiplets). This is strictly true as far as the metric of the scalar
manifold is strictly positive definite at the considered BPS critical
points. Indeed, by using special K\"{a}hler geometry one can prove the
following result, holding for any $\mathcal{N}=2$, $d=4$ supergravity \cite
{FGK} (such a result, \textit{mutatis mutandis}, holds also for $d=5$ \cite
{FG2}):
\begin{equation}
\left( D_{i}\overline{D}_{\overline{j}}V_{BH}\right) _{BPS}=2\left( g_{i%
\overline{j}}V_{BH}\right) _{BPS}.
\end{equation}

Reconsidering $\mathcal{N}=2$, $d=4$ supergravity, the Riemann tensor of the
special K\"{a}hler scalar manifold satisfies the following relation (see
\textit{e.g.} \cite{CDF-review} and Refs. therein)
\begin{equation}
R_{i\overline{j}l\overline{k}}=-g_{i\overline{j}}g_{l\overline{k}}-g_{i%
\overline{k}}g_{l\overline{j}}+C_{ilp}\overline{C}_{\overline{j}\overline{k}%
\overline{p}}g^{p\overline{p}},
\end{equation}
where the rank-$3$ completely symmetric tensor $C_{ijk}$ has the properties
\begin{equation}
\overline{D}_{\overline{l}}C_{ijk}=0,~D_{[l}C_{i]jk}=0.
\end{equation}
In particular, for homogeneous symmetric cubic special K\"{a}hler
geometries another set of relations holds \cite{GST2,CVP} (see also
\cite{ADFT}, \cite {DFT07-1} and Refs. therein; here and below
$z^{i}$ denote the complex scalars):
\begin{equation}
\begin{array}{l}
D_{l}C_{ijk}=0; \\
\\
C_{ijk}=e^{K}\partial _{i}\partial _{j}\partial _{k}f\left( z\right)
,~~f\left( z\right) \equiv \frac{1}{3!}d_{ijk}z^{i}z^{j}z^{k}; \\
\\
\overline{E}_{\overline{i}ijpq}\equiv g^{k\overline{k}}g^{r\overline{j}%
}C_{r(pq}C_{ij)k}\overline{C}_{\overline{k}\overline{i}\overline{j}}-\frac{4%
}{3}g_{\left( q\right| \overline{i}}C_{\left| ijp\right) }=0; \\
\\
d_{ABC}d^{B(PQ}d^{LM)C}=\frac{4}{3}\delta _{A}^{(P}d^{QLM)}.
\end{array}
\end{equation}

The $\mathcal{N}=2$, $d=4$ attractor equations read \cite{FGK}
\begin{equation}
2\overline{Z}D_{i}Z+iC_{ijk}g^{j\overline{j}}g^{k\overline{k}}\left(
\overline{D}_{\overline{j}}\overline{Z}\right) \overline{D}_{\overline{k}}%
\overline{Z}=0,  \label{N=2-attractors}
\end{equation}
$Z$ denoting the $\mathcal{N}=2$ covariantly holomorphic \textit{central
charge function} \cite{CDF-review}
\begin{equation}
Z\equiv e^{K/2}\left( X^{\Lambda }q_{\Lambda }-F_{\Lambda }p^{\Lambda
}\right) .
\end{equation}
Eqs. (\ref{N=2-attractors}) yield three classes of solutions \cite{BFM,BFGM1}%
:

\textit{i}) $\frac{1}{2}$-BPS solutions \cite{FGK}, with $Z\neq 0$ and $%
D_{i}Z=0$ $\forall i$. They saturate the BPS bound \cite{BPS}:
\begin{equation}
M_{ADM,BPS}^{2}=\left| Z\right| _{BPS}^{2},
\end{equation}
$M_{ADM}$ being the Arnowitt-Deser-Misner BH mass \cite{ADM}.

\textit{ii}) non-BPS solutions with $Z\neq 0$ and $D_{i}Z\neq 0$ for at
least some $i$ \cite{FGK,GIJT,K1,TT,BFM}. They do not preserve any
supersymmetry and do not saturate the BPS bound; indeed, for symmetric
spaces it holds that \cite{BFGM1}:
\begin{equation}
M_{ADM,non-BPS,Z\neq 0}^{2}=4\left| Z\right| _{non-BPS,Z\neq 0}^{2}>\left|
Z\right| _{non-BPS,Z\neq 0}^{2}.
\end{equation}
Such a result actually holds for homogeneous non-symmetric \cite{DFT07-1}
and also for generic cubic (at least within some particular assumptions \cite
{TT}) special K\"{a}hler geometries.

\textit{iii}) non-BPS solutions with $Z=0$ and $D_{i}Z\neq 0$ for at least
some $i$ \cite{AoB-book,BFGM1,DFT07-1,BMOS-1}. They do not preserve any
supersymmetry and do not saturate the BPS bound:
\begin{equation}
M_{ADM,non-BPS,Z=0}^{2}=\left[ g^{i\overline{j}}\left( D_{i}Z\right)
\overline{D}_{\overline{j}}\overline{Z}\right] _{non-BPS,Z=0}>0.
\end{equation}

As mentioned above, $\frac{1}{2}$-BPS critical points are stable; they have
no massless Hessian modes at all, and thus they do not have any associated
moduli space. The moduli spaces associated to the $\mathcal{N}=2$, $d=4$
non-BPS solutions with $Z\neq 0$ and $Z=0$ and to the $\mathcal{N}=2$, $d=5$
non-BPS solutions have been recently determined in \cite{Ferrara-Marrani-2}
(see also \cite{CFM1}); they are respectively given by Tables 2, 3 and 4 of
\cite{Ferrara-Marrani-2}.

As obtained in \cite{TT}, the $2n_{V}\times 2n_{V}$ (real form of the)
Hessian matrix of $V_{BH}$ at its non-BPS $Z\neq 0$ critical points in a
generic cubic special K\"{a}hler geometry of complex dimension $%
dim_{C}=n_{V} $ has $n_{V}+1$ strictly positive and $n_{V}-1$
vanishing eigenvalues. As pointed out above, in the homogeneous (not
necessarily symmetric) case, these latter $n_{V}-1$ massless Hessian
modes actually correspond to $n_{V}-1$ flat directions of
$V_{BH,non-BPS,Z\neq 0}$ \cite {Ferrara-Marrani-2}.

The same result holds also for generic cubic special K\"{a}hler
geometries, at least for some particular BH charge configurations \cite{CFM1}. This is simply seen \textit{e.g.} by splitting the complex scalars as $%
z^{i}=x^{i}-i\lambda ^{i}$, and considering the peculiar non-BPS $Z\neq0$-supporting BH charge configuration $%
Q_{0}^{T}=\left( p^{0},0,q_{0},0\right) $, for which the criticality conditions $%
\frac{\partial V_{BH}}{\partial x^{i}}=0$ can be solved by putting
$x^{i}=0$ $\forall i$. For such a case, in \cite{CFM1} $V_{BH}$ was
shown to acquire the following simple form:
\begin{equation}
\left. V_{BH}\right| _{x^{i}=0~\forall i,~Q=Q_{0}}=\frac{1}{2}\left[ \left(
p^{0}\right) ^{2}\mathcal{V}+\left( q_{0}\right) ^{2}\mathcal{V}^{-1}\right]
\equiv V_{BH}^{\ast }\left( \mathcal{V},p^{0},q_{0}\right) ,
\end{equation}
where $\mathcal{V}\equiv\frac{1}{3!}d_{ijk}\lambda ^{i}\lambda
^{j}\lambda ^{k}$. By rescaling $\lambda ^{i}\equiv
\mathcal{V}^{1/3}\widehat{\lambda }^{i}$,
it is immediate to realize that $V_{BH}^{\ast }\left( \mathcal{V}%
,p^{0},q_{0}\right) $ \textit{does not depend on any of the }$\widehat{%
\lambda }^{i}$. By definition, the $\widehat{\lambda }^{i}$s belong
to the geometrical \textit{locus} $\frac{1}{3!}d_{ijk}\widehat{\lambda }^{i}%
\widehat{\lambda }^{j}\widehat{\lambda }^{k}=1$; thus, they parameterize $%
n_{V}-1$ ``flat'' directions of $ \left. V_{BH}\right|
_{x^{i}=0~\forall i}$ at its non-BPS $Z\neq0$ critical points
supported by the charge configuration $Q_{0}$. Such $n_{V}-1$
``flat'' directions turn out to span nothing but the
$\left(n_{V}-1\right) $-dim. real special scalar manifold of the
corresponding $\mathcal{N}=2$, $d=5$ parent supergravity theory
\cite{CFM1}.

Let us now consider an explicit example, namely the \textit{magic} $\mathcal{%
N}=2$, $d=4$ supergravity theory based on the exceptional Jordan algebra $%
J_{3}^{O}$ over the octonions (see \textit{e.g.} \cite
{BFGM1,Ferrara-Gimon,Ferrara-Marrani-1,Ferrara-Marrani-2} and Refs.
therein). It is based
on the rank-$3$ homogeneous symmetric special K\"{a}hler manifold $\frac{%
E_{7(-25)}}{E_{6(-78)}\otimes U(1)}$ with $dim_{C}=n_{V}=27$; the charge
vector $Q$ sits in the fundamental representation $\mathbf{56}$ of $%
E_{7(-25)}$.

The $\frac{1}{2}$-BPS attractors are supported by a $Q$ belonging to the BPS
orbit $\frac{E_{7(-25)}}{E_{6(-78)}}$ ($dim_{R}=55$); due to the compactness
of $E_{6(-78)}$, there is no BPS moduli space at all.

On the other hand, the non-BPS $Z\neq 0$ attractors are supported by a $Q$
belonging to the $55$-dim. non-BPS orbit $\frac{E_{7(-25)}}{E_{6(-26)}}$, $%
E_{6(-26)}$ being a non-compact real form of the exceptional group $E_{6}$;
the corresponding non-BPS $Z\neq 0$ moduli space reads $\frac{E_{6(-26)}}{%
F_{4(-52)}}$ ($dim_{R}=26$), where $F_{4(-52)}=m.c.s.\left(
E_{6(-26)}\right) $ \cite{Gilmore}. It is nothing but the rank-$2$, real
special scalar manifold of the corresponding $\mathcal{N}=2$, $d=5$ parent
supergravity theory \cite{CFM1}.

The non-BPS $Z=0$ attractors are supported by a $Q$ belonging to the $55$%
-dim. non-BPS orbit $\frac{E_{7(-25)}}{E_{6(-14)}}$, $E_{6(-14)}$
being the\ only other non-compact real form of $E_{6}$\ contained in
$E_{7(-25)}$ \cite {Gilmore}; the corresponding non-BPS $Z=0$ moduli
space is the rank-$2$, homogeneous symmetric (not
special) K\"{a}hler manifold $\frac{E_{6(-14)}}{%
SO(10)\otimes U(1)}$ ($dim_{C}=16$) \cite{CFM1}, where
$SO(10)\otimes U(1)=m.c.s.\left( E_{6(-14)}\right) $ \cite{Gilmore}.

The corresponding parent theory in $d=5$ is the \textit{magic} $\mathcal{N}%
=2 $, $d=5$ supergravity over $J_{3}^{O}$ (see \textit{e.g.} \cite
{GST2,FG,Ferrara-Marrani-2} and Refs. therein). For such a theory,
the BPS charge orbit coincides with $\frac{E_{6(-26)}}{F_{4(-52)}}$
itself \cite{FG}, and there are no BPS massless Hessian modes
\cite{FG2}. The unique class of
non-BPS attractors with non-vanishing cubic invariant $I_{3}$ (see \textit{%
e.g.} \cite{FG2} and Refs. therein) is supported by the $26$-dim. BH charge
orbit $\frac{E_{6(-26)}}{F_{4(-20)}}$, $F_{4(-20)}$ being the only\
non-compact real form of the exceptional group $F_{4}$ contained in $%
E_{6(-26)}$ \cite{Gilmore}. The corresponding non-BPS moduli space is the
rank-$1$, homogeneous symmetric manifold $\frac{F_{4(-20)}}{SO(9)}$ ($%
dim_{R}=16$) \cite{Ferrara-Marrani-2}, where $SO(9)=m.c.s.\left(
F_{4(-20)}\right) $ \cite{Gilmore}.

Finally, it is worth remarking that the non-BPS $d=5$ attractors can
give rise to both $Z\neq 0$ and $Z=0$ non-BPS $d=4$ critical points,
depending on the sign of an extra Kaluza-Klein charge \cite{CFM1}.
This implies that the moduli space of non-BPS $d=5$ attractors is
contained in the moduli spaces of both species ($Z\neq 0$ and $Z=0$)
of non-BPS $d=4$ attractors, as pointed out in \cite{CFM1} (and as
given by the Tables 2, 3 and 4 of \cite {Ferrara-Marrani-2}).

\section*{\textbf{Acknowledgments}}
The topics covered in this review are mainly based on collaborations
with: L. Andrianopoli, S. Bellucci, A. Ceresole, R. D'Auria, G.
Gibbons, E. Gimon, M. G\"{u}naydin, R. Kallosh, A. Strominger, M.
Trigiante.

 The work of S.F. has been
supported in part by European Community Human Potential Program
under contract MRTN-CT-2004-005104 \textit{``Constituents,
fundamental forces and symmetries of the universe''} and the
contract MRTN-CT-2004-503369 \textit{``The quest for unification:
Theory Confronts Experiments''}, in association with INFN Frascati
National Laboratories and by D.O.E. grant DE-FG03-91ER40662, Task C.

The work of A.M. has been supported by a Junior Grant of the \textit{%
``Enrico Fermi''} Center, Rome, in association with INFN Frascati
National Laboratories.

\end{document}